\magnification=1200
\baselineskip=0.8truecm
\centerline{\bf Scaling Regimes in the Distribution of Galaxies}
\bigskip

\centerline{G. Murante, A. Provenzale,}
\centerline{Istituto di Cosmogeofisica, Corso Fiume 4, I-10133 Turin}
\medskip
\centerline{E. A. Spiegel,}
\centerline{Dept. of Astronomy, Columbia University, New York, NY 10027, USA
}
\medskip
\centerline{R. Thieberger,}
\centerline{Dept. of Physics, Ben Gurion University, Beer Sheva, Israel}

\bigskip
\bigskip

\noindent{\bf Abstract}
\noindent
If we treat the galaxies in published redshift catalogues as point sets,
we may determine the generalized dimensions of such sets by standard
means, outlined here.   For galaxy separations up to about 5 Mpc, we find
the dimensions of the galaxy set to be about 1.2, with not a strong
indication of multifractality.  For larger scales, out to about 30 Mpc,
there is also good scaling with a dimension of about 1.8.  For even
larger scales, the data seem too sparse to be conclusive, but we find
that the dimension is climbing as the scales increase.  We report
simulations that suggest a rationalization of such measurements, namely
that in the intermediate range the scaling behavior is dominated by flat
structures (pancakes) and that the results on the smallest scales are a
reflection of the formation of density singularities.

\bigskip
\bigskip \noindent
{\bf 1. Introduction}

\noindent
An interesting question in modern cosmology is how to reconcile the
very inhomogeneous distribution of galaxies with the very isotropic
distribution of the cosmic background radiation.  The conventional
picture is that the marginally detectable anisotropies of the radiation
field are the seeds of the impressive structures seen in the galaxy
distribution.   The transition from weak initial perturbation to
strong clustering is supposed to be initiated by gravitational
instability.  This process, which is weak in an expanding medium, is
nevertheless called upon to take the perturbations into a nonlinear
collapse phase that produces well defined structures that are reflected
in the lumpy distribution of galaxies.

In the present work, we try to make sense of the complicated galaxy
distribution in terms of simple --- perhaps overly simple --- pictures.
We are suggesting that the data on the galaxy distribution imply that
there are two principal scales of structure formation each with its own
characteristics.  The aim of this informal general discussion is explain
why we believe that we can detect these regimes in the analysis of
publically available galaxy catalogues.  The existence of a possible third
regime at the largest scales may also be discernable.  The final
clarification of course awaits the coming of the tidal wave of data that
should soon engulf us and which may perhaps wash away all our present
conceptions.

Before we describe how the galaxy distribution is quantified and the
regimes are detected, we briefly describe the simulations of gravitational
clustering that have produced the preconceptions that we bring to
the analysis.  Then we outline one of the main methods of quantification of
structure that has been used in this problem.  Finally, we offer a
preliminary interpretation of the observed structures.
\bigskip\noindent
{\bf 2. Density singularities}

\noindent
Gravitational instability is usually assumed to be the primary
mechanism that shapes the galaxy distribution.  There may be other
interactions that play a role in the process,  depending on the era
and on what matter and fields are present, but we ignore those here.
Most of the models on the market are based on a strong preponderance of
dark matter, and this is consistent with many lines of evidence.  It is
not clear what this matter is, but we shall here suppose that it is rather
cool and describe a CDM (cold dark matter) scenario.

Early analytic studies have indicated that gravitational collapse may
generate virialized mass concentrations, with approximately isothermal
density profiles (Gunn and Gott 1972, Fillmore and Goldreich
1984, Bertschinger 1985, Gurevich and Zybin 1988), with details depending
on the initial density perturbations and on the value of $\Omega$.  A
study by Henriksen and Widrow (1995) based on Vlasov equation in expanding
space produces local singular density distributions with density
proportional to $r^{-\alpha}$ where $1.5 \le \alpha \le 3$ and $r$ is
the distance from the singularity.

The means for simulating structure formation are diverse. Some
effort has  gone into modeling with the Vlasov equations complemented by
the Poisson equation.   This latter would in principle  be a good way to
proceed if one could really carry it out. In fact, it is difficult to
follow in detail the development of plasma turbulence through the
collective effects that it contains..  Such collective effects perhaps make
the gravitational problem more like a fluid dynamical problem than
a particle dynamics problem.

More typically, numerical simulation of the structure formation process
has been pursued by N-body techniques, where point masses, representing
fluid elements, move under the influence of their own gravitational
interaction in an expanding background. On the scales of galaxies and
galaxy clusters, these simulations generally confirm the presence of
strong density concentrations with a local density growing algebraically,
at least in a certain range of scales (Navarro, Frenk and White 1996, 1997).

Many simulations have been performed and we may mention the recent report
of large scale calculations of this kind by Zurek and Warren (1994).
Our own simulations, which we describe in this section, were performed
with an adaptation of Couchman's (1991) public AP3M code for present
purposes.  These simulations have $N_p=128^3$ massive particles on a
Cartesian grid of $N_g=128^3$.  For the initial conditions, we introduce,
as is usual, a small perturbation superposed onto a homogeneous density
field that is a solution to the Friedmann-Lemaitre equations. The average
density has been taken to be equal to the critical density $\rho_{cr}$,
so that $\Omega= \rho_{av}/\rho_{cr}=1$; this makes it easier for
structures to form than do the observationally suggested values which are
noticeably less than unity.  The initial conditions have Gaussian,
scale-free density perturbation spectra with random Fourier phases and
density power spectrum
$$ P(k)\propto k^n \ . \eqno(2.1) $$

During the computed evolution, strong density concentrations form.
For each particle cluster, the center of mass has been determined and the
radial density profile has been evaluated.  Each profile has been
normalized to the half-mass radius (the radius within which
half of the mass in a cluster is containted) and the average density
profile has been calculated.

Figure 1 shows the averaged dark halo
profiles for the four initial conditions
considered, at the time when the variance of the density field,
coarse-grained on a scale of $1/16$ of the domain size is unity.
At scales smaller than the half-mass radii, the dark halos have
power-law profiles for all the four initial
conditions considered. These profiles are slightly steeper for larger
$n$, showing also a tendency to display a larger range of power-law
behavior for larger values of $n$.

What is the reason for the difference between the four average
profiles, and what is their time evolution? Are they
evolving in time or do they have a logarithmic slope which is
constant in time?

Figure 2 shows the average profiles of collapsed objects, obtained
from the three initial conditions with $n=1,0,-1$, at different
times. This figure shows that, independently of the initial
conditions, the slope of the profile is a function of the
central density of the collapsed object, the profile being
steeper for larger central density. The average profiles of
objects with similar central density, produced by simulations with
different initial conditions, are the same. For the
three spectral indices shown in Figure 3,
the profile steepens, and comes closer to a power-law,
as time increases.

A different behavior is observed for $n=-2$. In this case, shown in
Figure 3, the profiles at different times do not evolve; rather, they
keep the same shape and logarithmic slope.  The initial conditions for
the CDM scenario are often assumed to have a power spectrum with $n\approx
-2$ on small scales.  This suggests that dark halos for CDM initial
conditions do not evolve in time, once the transients have decayed.

We turn now to a discussion of the evolved structures.  In doing
this we must leave several questions unanswered: why do the
$n=-2$ initial conditions behave so differently from the other
three cases considered? For $n=1,0,-1$, do the dark halo profiles
evolve in time forever, with an ever-increasing steepness, or do they
reach a limiting slope after which they do not evolve further? To answer
these questions, one should properly understand the behavior of
gravitational collapse at small scales. On more general grounds, we
recall that the results reported here (and in several other
numerical studies) refer to cold initial conditions. Does the addition of
random dispersion velocities significantly modify the picture? And,
finally, what happens when one includes hydrodynamical effects, such that
the collapse is not purely gravitational any longer?

\bigskip\noindent
{\bf 3. Quantifying Cosmic Scaling}

\noindent
Early analyses of the galaxy distribution were based upon the
the correlation function (Groth and Peebles 1977,
Peebles 1980) but, in recent years, the
correlation integral (Grassberger and Procaccia 1983, Paladin and
Vulpiani 1984, Borgani 1995) has also been used.  It seems
that if the galaxy distribution is a fractal, the latter is advantageous
since the definition of the correlation involves the density of galaxies.
Though the notion of density is intuitively reasonable, it is not well
defined for a fractal (Coleman, Pietronero and Sanders 1988, Coleman and
Pietronero 1992) and this makes the use of correlations problematic.  At
any rate, in the study of analytically understood point sets, the
correlation integral is a more reliable tool of analysis
(Thieberger, Spiegel and Smith 1990), so it seems worthwhile to describe it
briefly here.  We treat the galaxies as a point set for this purpose.

Let ${\cal N}_i(r)$, be the number of galaxies of the set lying
within a distance $r$ of the typical (the i$^{th}$) galaxy.  This number
is then averaged over the galaxies in the set to give us the number of
galaxies within a distance $r$ of a typical galaxy.  If there are a lot of
galaxies around, we may replace the sum in the calculation of ${\cal N}$
by an integral and write
$$
{\cal N }(r) = 4\pi \int_0^r n(s) s^2 ds \ .\eqno(3.1) $$
The radial distribution function $n$ is not a good density
of galaxies since it does not in general tend to a finite limit at
extremes of $r$.

If the galaxies in the set are found in a finite volume, we may divide
the total number of them by this volume and call this $n_0$.  The
correlation used in the study of the galaxy distribution may then be
written as $\xi=n/n_0-1$.  (The choice of $n_0$ is not crucial and it
may be defined otherwise than here.)  For a uniform Poisson distribution,
$\xi=0$, the galaxies are uncorrelated and $n=n_0$.  If $\xi$ is
positive, the galaxy positions are correlated.

When $r$ goes to zero, so must ${\cal N}(r)$ and the simplest way
to achieve this is to let $n$ go as a power, say,
$n = n_0(r/r_0)^{-\gamma}$, where $r_0$ is a
conveniently chosen length.  For small $r/r_0$, we see that
$\xi \propto r^{-\gamma}$, so there seems little to choose between
the two approaches.  But when the data are few, the tendency is
to let $r/r_0$ get close to unity, and the differences appear in the
results.  Enough has been written about these differences so that we have
simply chosen one approach, which we now outline.

The correlation integral (Grassberger and Procaccia 1983) can be defined
as $$ C_2(r)= {1\over {N(N-1)}}\Sigma_i \Sigma_{j\ne i} \Theta(r - |{\bf
X}_i - {\bf X}_j|) \eqno(3.2)$$ where $\Theta$ is the Heaviside function.
The summations are over the whole set of $N-1$ galaxies with coordinates
${\bf X}_j$, $j\ne i$.  (Modifications of this formula to allow for
the effects of the finiteness of the sample have been used in dealing
with the data. We omit these technicalities here but see Murante {\it et
al.} (1997).) We see from the formula that $C_2(r)={\cal N}(r)/N$.  The
quantity $C_2(r)$ is thus proportional to the volume integral of $n(r)$.
As $r$ gets small, $C_2$ must go to zero and for general
distributions, in the limit $r \to 0$, and we express this as
$C_2 \propto r^{D_2}$.  The exponent $D_2$ is called the correlation
dimension and we see that $D_2=3-\gamma$  (Provenzale 1991).

As with the correlations, it is possible study higher order statistics.
To generalize the correlation integral formalism (Grassberger and
Procaccia 1983) one introduces correlation integrals defined as
$$ C_q(r)= \left({1\over {N}}\Sigma_i \left[
{1\over {N-1}} \Sigma_{j\ne i} \Theta(r - |{\bf X}_i - {\bf
X}_j|)\right]^{q-1}\right)^{1\over q-1} \eqno(3.3)$$
where $q$ is a parameter that defines the order of the moment.
For $q=2$ a two-point probability (second order moment) is evaluated and
the standard correlation integral is recovered.  More generally, for any
integer value of $q$, $C_q(r)$ is the fraction of
of $q$-tuples in the set whose members lie within a distance $r$ of one
another.  For sufficienly small $r$, $C_q$ will go to zero for $q > 1$
and, for a typically well-behaved set, it will vanish like $r^{D_q}$.
The index $D_q$ is called a generalized or Renyi dimension (Renyi
1970, Halsey {\it et al.} 1986).   A fractal for which the dimensions are
all the same ($D_q$ independent of $q$) is called a homogeneous fractal,
or a monofractal.  The more general cases with $D_q$ depending on $q$ are
called multifractals.  Some authors reserve the general term fractal
for the special case of a monofractal, but here we retain the general
sense of the term fractal, with the multifractal as a particular case.
\goodbreak
\bigskip\noindent
{\bf 4. Lacunarity}

\noindent Another feature of the scaling analysis may ultimately prove
interesting for the study of the galaxy distribution
once the data are sufficiently abundant.  As yet we have only some
preliminary results on this, but we find them intriguing
enough to mention here.

For the typical case, the correlation integrals go to zero like a power
of the separation.  This behavior may be considered as the first term in
an asymptotic expansion of $\log C_q$
in $\log r$.  More generally, we may seek an expansion of the form $$
\log C_q(\log r) = D_q \log r + E_q + {F_q \over \log r} + ...\ .
\eqno(4.1) $$

Higher terms are hard to detect when there are few data, but it may be
possible to get an estimate of the second term.  When only two terms
are kept in the expansion, we have the representation $$
C_q=\Lambda_q r^{D_q} \ . \eqno(4.2)
$$  Mandelbrot (1982) has named $\Lambda$ the lacunarity.

A weak dependence of $E_q$ on $r$ is compatible with this representation.
This is seen easily for the case of the homogeneous fractal, for which
the generalized dimensions, $D_q$, are all the same, $D_q=D$, say.  We
may expect in that case that any statistical moment $C(r)$ satisfies the
scaling law
$$ C(r) = a C(br) \eqno(4.3) $$
with constant $a$ and $b$.
This functional equation has solutions of the form (4.2) with
$$
D_q \equiv D={\log a\over \log b}.
\eqno(4.4) $$

As we see, (4.2) and (4.4) satisfy the functional equation (4.2) even when
$\Lambda$ depends on $\log r$ provided that $\Lambda(\log r - log
b)=\Lambda(\log r)$.  In this case, we call $\Lambda$ the
lacunarity function, or LF, and we may identify $\log \Lambda$ with $E_q$
in the asymptotic development, so long as $E_q$ remains of order unity.
The appearance of a periodic dependence of $E_q$ on $\log r$ is typically
seen in monofractals, when there are enough data.  Solis and Tao (1997)
have seen this in theoretical multifractals, though more weakly.  In the
case of the theoretical fractals, the period of the LF is a remnant of the
decimation procedure that produced the fractal.

\bigskip\noindent
{\bf 5. Cosmic Scaling Regimes}

\noindent
We concentrate here on the analysis of redshift catalogues, so we shall
not go into the issues that the study of position catalogues entail such
as the effect of projection of the spatial distribution onto the celestial
sphere.  A full discussion of this issue would involve the basic notions
developed in the study of stellar statistics and the analysis of radio
source counts.  In the latter, a dependence on $r$ was often overlooked
and anomalies in the data were rationalized by assuming a dependence on
time.  We shall not go that far back into the history of the subject.

The first analyses of the distribution of the galaxies on
the celestial sphere to gain wide acceptance were based on the correlation
function and they found an approximate scaling regime on scales
smaller than about 5 $h^{-1}$ Mpc, where $h$ is the Hubble constant in
units of 100 km/sec Mpc$^{-1}$ (Groth and Peebles 1977,
Peebles 1980). Out to these scales, the
two-point galaxy correlation function has an approximate power-law shape
with $\xi(r) \propto r^{-\gamma}$ where $\gamma \approx 1.8$ (Peebles
1980).  In the fractal vernacular, this would mean that the correlation
dimension of the galaxy distribution is $D_2 \approx 1.2$ out to an outer
scale of 5 $h^{-1}$ Mpc.

When the data were later analyzed using correlation integrals, the result
$D_2 \approx 1.8$ was found (Thieberger, Spiegel and Smith 1990,
Martinez and Jones 1990,
Borgani 1995).  The explanation for this discrepancy with the
earlier work, at least the one that we adopt, is that the scaling
law on scales under 5 Mpc is different
from those at larger scales (Guzzo {\it et al.}
1991, Murante {\it et al.} 1996).  That is, we
have $0.8 < D_q < 1.4$ for $q \ge 2$ at scales below 5 $h^{-1}$ Mpc,
whereas, on intermediate scales, the value of $D_2$ is just under $2$;
we find $1.8$.  On the very largest scales, above about $30Mpc$,
we find that $D_2$ has begun another rise.

To understand the meaning of $D_2 = 3- \gamma \approx 1.2$ for scales
less than 5 Mpc, we  have computed the dimension of a random
distribution of density singularities with power law density
distributions.  Once strong clustering has occured, simulations of the
kind we discussed in the section 2 suggest that the collapse into
clusters continues into the formation of local structures with approximate
power-law density profile around the center of the collapsed distribution.
Within such distributions, there is no fractal structure in the sense that
this term is now normally used, though one may broaden that sense.
When there are local smooth structures, the usual notion of density
may be used, even if its definition is still not so clear.  As mentioned
in the discussion on dynamics, roughly spherical structures with density
distributions with densities varying like $r^{-\alpha}$ form with $r$
being the distance from the sinularity center.  The value $\alpha=2$,
which is appropriate for the isothermal sphere, is just the one that
produces flat rotation curves in a disk around the singularity center.  We
have found that a random distribution of such power-law density
singularities can give global scaling properties like those seen on scales
seen for sizes under about five Mpc, for a density exponent, $\alpha$,
of order two.

On the other hand, simulations show that the first structures to form
are highly flattened, as anticipated by Zeldovich (1970).  Such structures
produce a fractal dimension of about 2.  The natural interpretation
is that the initial formation of pancakes of this kind then leads to the
formation of substructures which collapse down to singularities and the
two separate regimes produce the two observed correlation dimensions.
The value of $D_2 \approx 1.8$ that now seems to hold in the scale range
of 5-50 Mpc is compatible with the pancake regime.

As we go to larger scales, $D_2$ seems to be increasing, but the data
are too sparse as yet for a definite conclusion.  Many believe that there
has not been enough time as yet to have allowed the formation of well
defined structures on scales of hundreds of megaparsecs.  On the other
hand, Pietronero and collaborators defend the de Vaucouleur idea that
there is structure at all scales in terms of these same data.  That vision
may call for a `spooky action at a distance' in Einstein's famous phrase
and so test our notions of causality.

In conclusion, we believe that the considerations we have
outlined here, rationalize the observed exponents but they do not
tell us why the scaling regimes are what is observed.  One way to go
further would be to look at finer details.  To do this, we have made an
attempt to detect the lacunarity function of the galaxy distribution.  For
this purpose, we have used a sample of 30000 galaxies from the CfA-ZCAT
galaxy catalogue (Huchra {\it et al.} 1993).  The reductions are described
elsewhere (Provenzale, Spiegel and Thieberger 1997).  A plot of $$
\log  \Lambda = \log C_q - D_q \log r \eqno(5.1) $$
vs. $\log r$ for $q =2,3,4$ is shown for the northern galaxies,
see Figure 4.
We see perhaps one period of the anticipated oscillation.

Since the amplitude of the oscillation in the LF is so small, we must
of course be wary in accepting it as real.  Nevertheless, there are some
features that do make it seem worth pursuing this matter further.  First,
there is the apparent constancy of what we may opitmistically call the
period for the three values of $q$ that we have examined.  This is
suggestive of monofractality.  Second, although the LF for the southern
hemisphere is not nearly so well defined as for the northern hemisphere,
its `period' is the same as in the northern hemisphere.  If this result
holds up, it is striking evidence of statistical homogeneity of the
galaxy distribution on large scales.  Since we do not see this kind of
oscillatory LF for the smallest scales the LF may also be in keeping with
our interpretation of the dimensions on those scales.

\vfill
\eject
\noindent{ \bf References.}

\noindent Bertschinger, E., 1985, Astrophys. J. Supp., 58, 39.

\noindent Borgani S., 1995, Phys. Rep., 251, 1.

\noindent Coleman P.H., Pietronero L. and Sanders R.H., 1988,
Astron. \& Astrophys., 200, L32.

\noindent Coleman P.H. and Pietronero L., 1992, Phys. Rep.,
231, 311.

\noindent Couchman, H.M.P., 1991, Astrophys. J. Lett., L386, 23-26.

\noindent Fillmore J.A. and Goldreich P., 1984, Astrophys. J., 281, 1.

\noindent Gurevich A. V. and Zybin P., 1988, Sov. Phys. JETP, 67, 1957.

\noindent Guzzo L., Iovino A., Chincarini G., Giovanelli R. and
Haynes M.P., 1991, Astrophys. J. Lett., 382, L5.

\noindent Grassberger P. and Procaccia I., 1983, Phys. Rev. Lett.,
50, 346.

\noindent Groth E.J. and Peebles P.J.E., 1977, Asrophys. J., 217, 385.

\noindent Gunn J. and Gott J.R., 1972, Astrophys. J.  281, 1.

\noindent Halsey T.C., Jensen M.H., Kadanoff L.P., Procaccia
I. and Shraiman B.I., 1986, Phys. Rev. A, 33, 1141.

\noindent Henriksen R.N. and Widrow L.M., 1995, Mon. Not. R. Astron.
Soc., 276, 679.

\noindent Huchra J.P., Geller M.J., Clemens C.M., Tokarz S.P. and Michel
A., 1993, CfA Redshift Catalogue; 10-3-93 version available at
ftp://cdsarc.u-strasbg.fr/pub/cats/VII/164.
\noindent Mandelbrot B.B., 1982, {\it The Fractal Geometry of Nature}
(San Francisco: Freeman).

\noindent Martinez V.J. and Jones B.J.T., 1990, Mon. Not. R.
Astron. Soc., 242, 517.

\noindent Murante G., Provenzale A., Borgani S., Campos A.
and Yepes G., 1996, Astroparticle Phys., 5, 53.

\noindent Murante G. Provenzale A. Spiegel E.A. and Thieberger R.,
1997, Mon.  Not.  R.  Astron.  Soc., 291, 585.

\noindent Navarro J.F., Frenk C.S. and White S.D.M., 1996,
Astrophys. J., 462, 563.

\noindent Navarro J.F., Frenk C.S. and White S.D.M., 1997,
Astrophys. J., 490, 493.

\noindent Paladin G. and Vulpiani A., 1984, Nuovo Cimento Lett., 41,
82.

\noindent Peebles P.J.E., 1980, {\it The Large Scale Structure of the
Universe} (Princeton: Princeton Univ. Press).

\noindent Provenzale A., 1991, in {\it Applying Fractals in Astronomy},
A. Heck and J.M. Perdang Eds. (Berlin: Springer).

\noindent Provenzale A., Spiegel E.A. and Thieberger R., 1997, CHAOS,
7, 82.

\noindent Renyi A., 1970, {\it Probability Theory} (Amsterdam: North-Holla=
nd).

\noindent Solis F.J. and Tao L., 1997, Phys. Lett. A, 228,
351.

\noindent Thieberger R., Spiegel E.A. and Smith L.A., 1990, in
{\it The Ubiquity of Chaos}, S. Krasner Ed. (AAAS Press).

\noindent Zeldovich Ya.B., 1970, Astrofizika(A) 6, 319.

\noindent Zurek W.H. and Warren M.S., 1994, Los Alamos Science, No. 22,
58.

\vfill
\eject
\noindent {\bf Figure Captions.}
\medskip
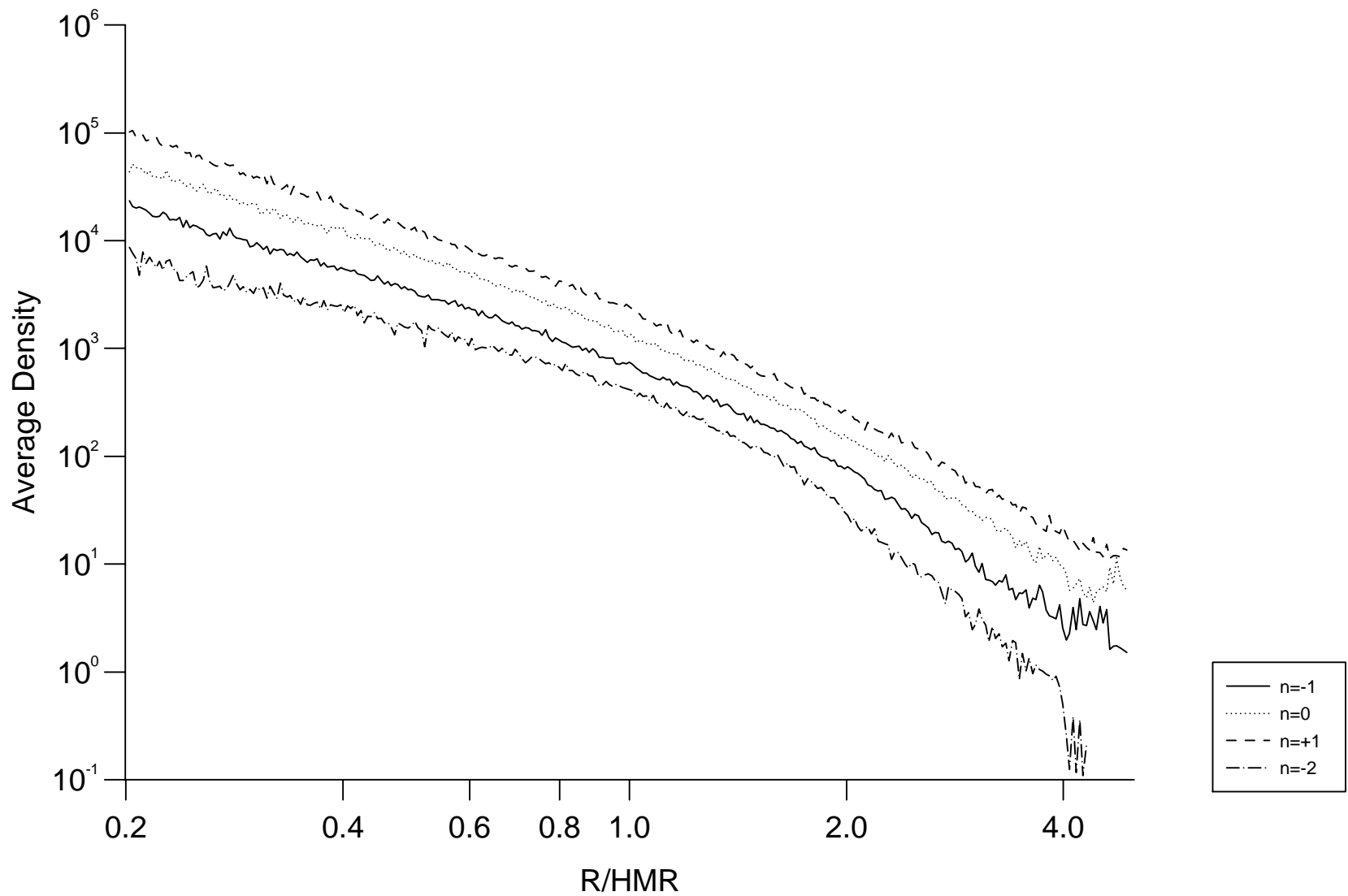
\noindent Figure 1.
\medskip
\noindent
Average density profiles of the dark halos
for the four scale-free initial conditions
considered in the text, at the time when the density
variance is 1 at 1/16th of the simulation box.
\medskip
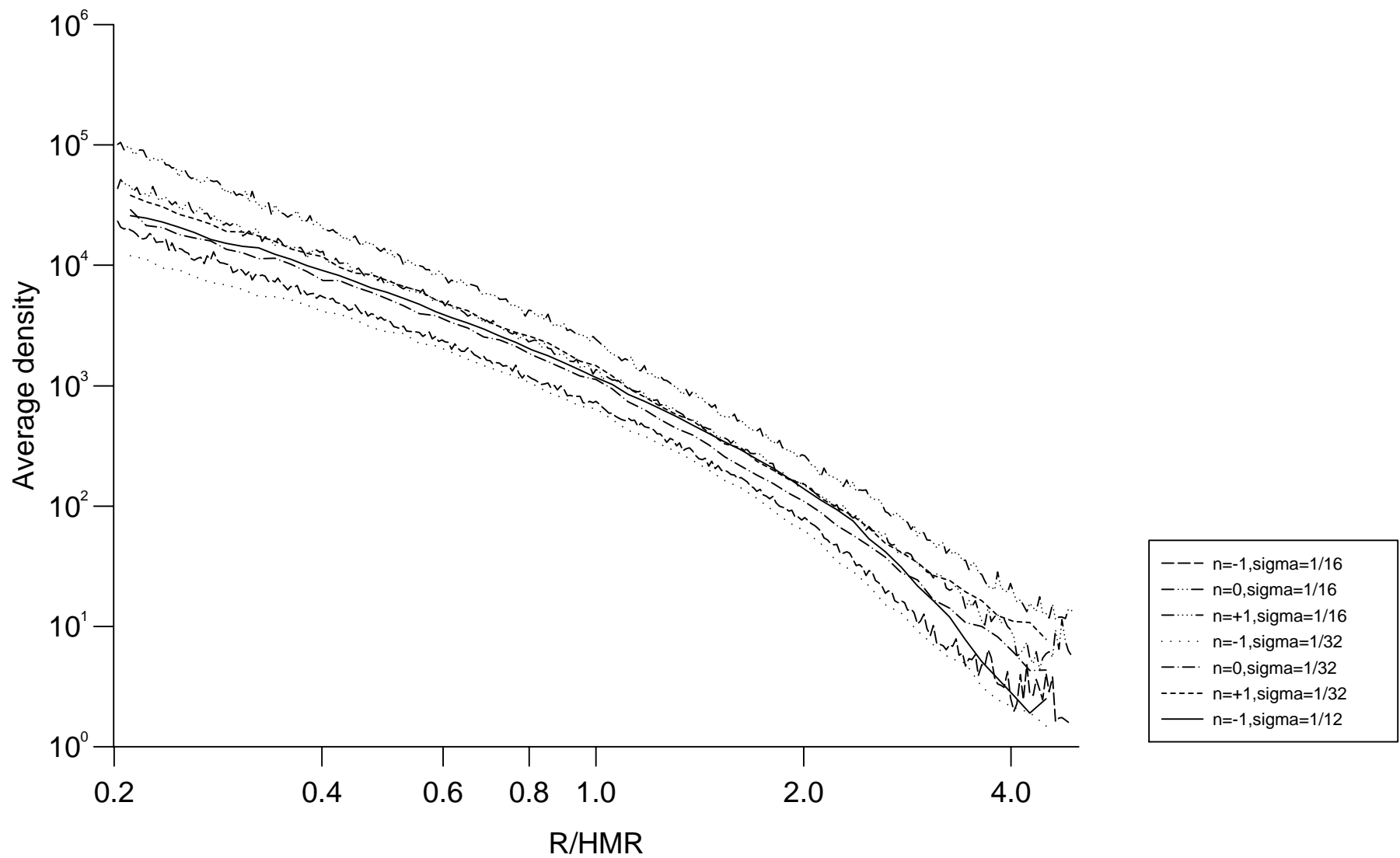
\noindent Figure 2.
\medskip
\noindent
Average profiles of the dark halos produced by cosmological N-body
simulations for
three scale-free initial conditions with spectral
indeces $n=1,0,-1$, at different
evolutive times.
\medskip
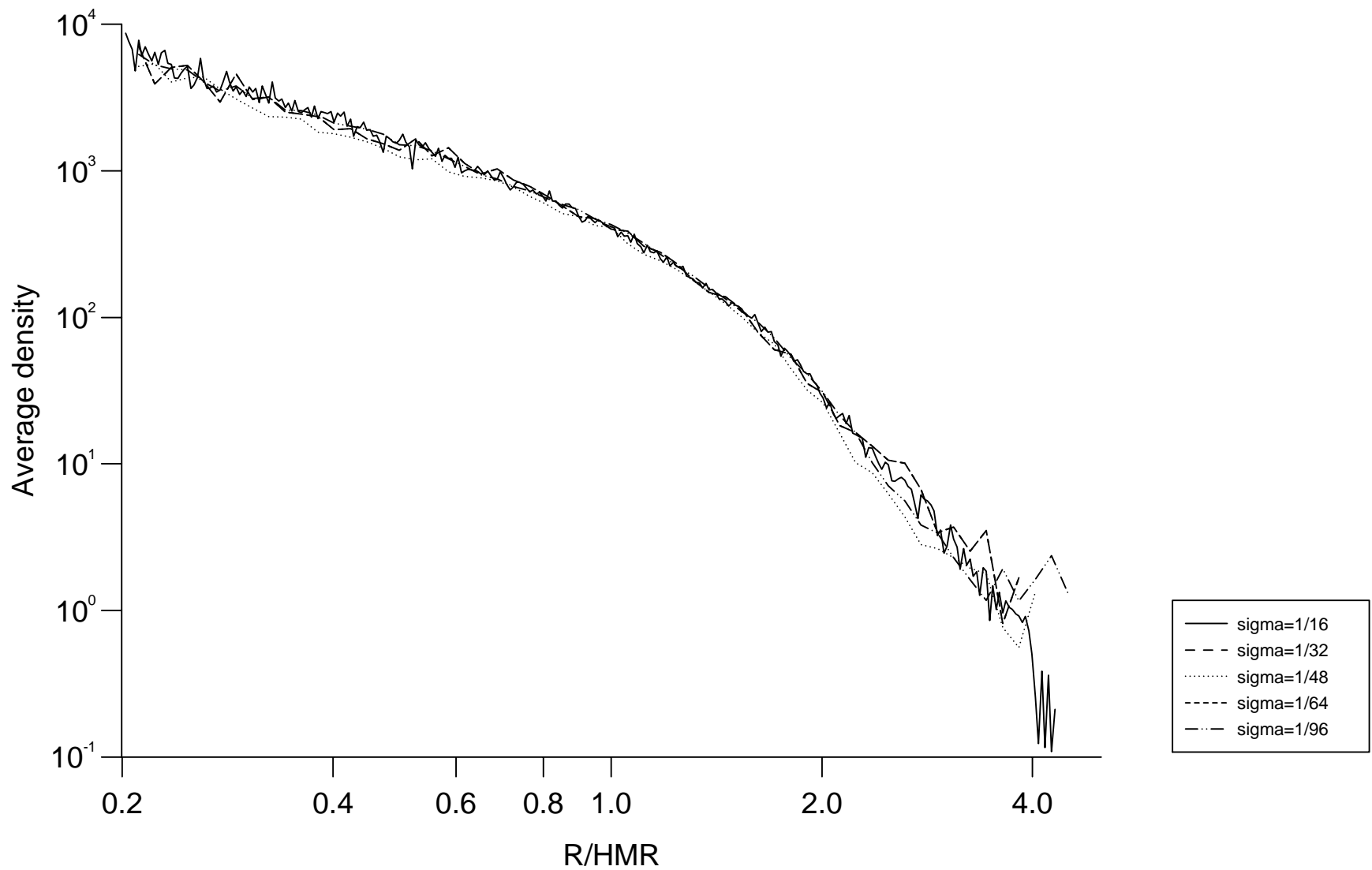
\noindent Figure 3.
\medskip
\noindent
Average profiles of the dark halos produced by a cosmological N-body
simulation with
scale-free initial conditions and spectral
index $n=-2$, at different
evolutive times.
\medskip
\noindent Figure 4.
\medskip
\noindent
Lacunarity function for $q=2,4,6$ for the CfA galaxies in the
Northern galactic emisphere.
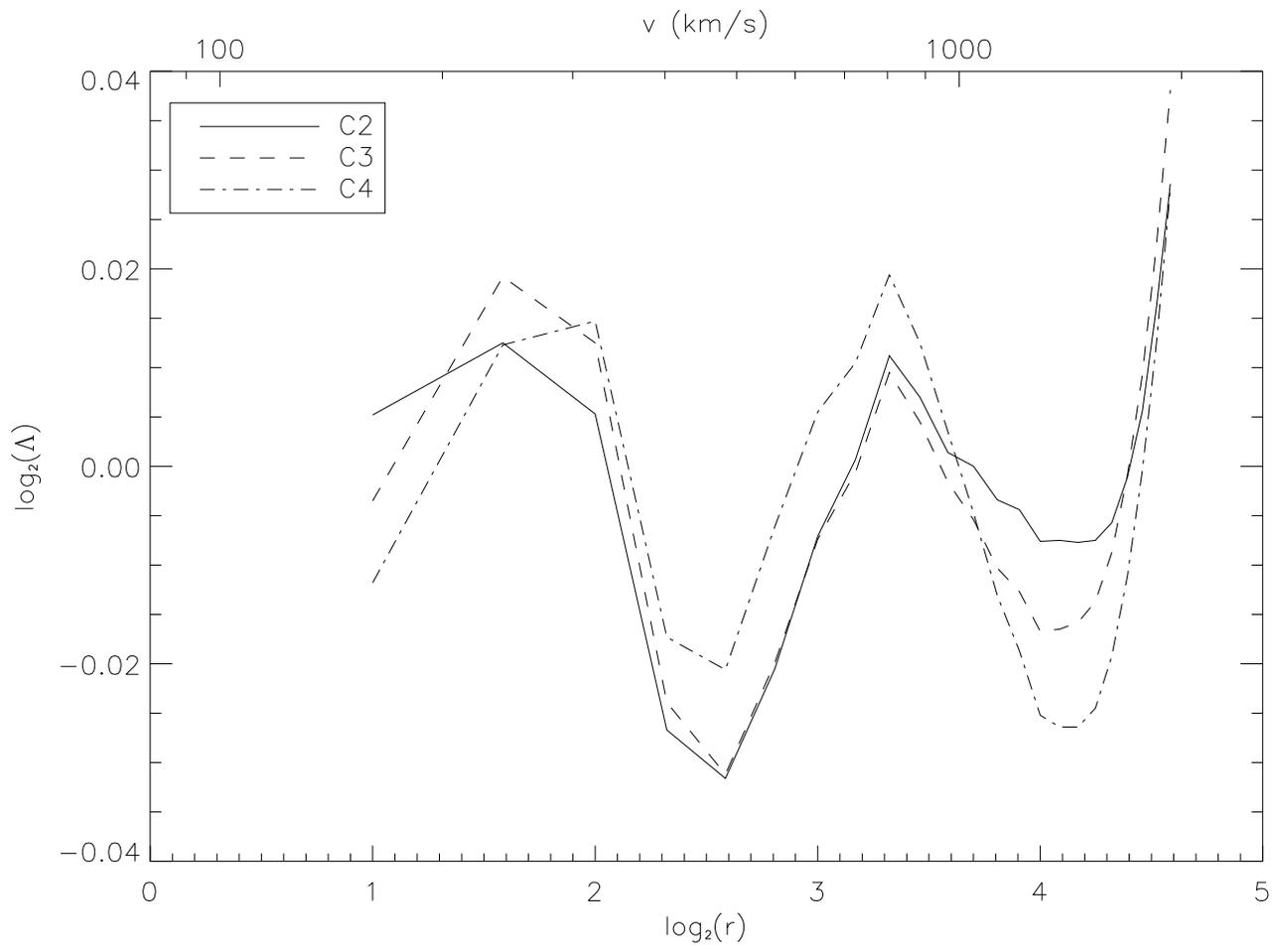
\end